\documentclass[preprint2,final]{aastex}
\shorttitle{Erosion of icy cores in gas giants}
\shortauthors{Wilson et al.}

\begin{document}
%\scriptsize

\title{Solubility of water ice in metallic hydrogen: consequences for core erosion in gas giant planets}

\author{H. F. Wilson}
\affil{Department of Earth and Planetary Science, University of California, Berkeley, CA 94720, USA.}
%\email{hubbard@lpl.arizona.edu}

 \and

\author{B. Militzer}
\affil{Departments of Earth and Planetary Science and of Astronomy, University of California, Berkeley, CA 94720, USA.}
%\email{militzer@berkeley.edu}

\begin{abstract}
Using \emph{ab initio} simulations we investigate whether water ice is
stable in the cores of giant planets, or whether it dissolves into the
layer of metallic hydrogen above. By Gibbs free energy calculations we
find that for pressures between 10 and 40 Mbar the ice-hydrogen
interface is thermodynamically unstable at temperatures above
approximately 3000~K, far below the temperature of the core-mantle
boundaries in Jupiter and Saturn. This implies that the dissolution of
core material into the fluid layers of giant planets of giant planets
is thermodynamically favoured, and that further modelling of the
extent of core erosion is warranted.
\end{abstract}


\section{Introduction}

According to core accretion models \citep{mizuno-78}, giant gas
planets such as Jupiter and Saturn formed via the accumulation of an
protocore of rock and ice which gained solid material until it reached
sufficient size to begin accreting the gaseous component of the
protosolar nebula. The existence of solid ice in the outer solar
system promotes the rapid growth of the more massive protocores
allowing the accretion of large quantities of gas necessary for
Jupiter-sized planets. Giant planets thus have a dense core of rock
and ice surrounded by a H-He envelope. It is not known, however,
whether the initial dense core remains stable following the accretion
of the H-He outer layer or whether the core erodes into the fluid
hydrogen-rich layers above \citep{stevenson-pss-82,guillot-book}.

The gravitational moments of Jupiter and Saturn, which have been
measured by prior planetary missions and will be determined for
Jupiter with unprecedented accuracy by the upcoming Juno mission, may
be used in combination with interior models
\citep{militzer-apj-08,guillot-book,hubbard-ass-05,saumon-apj-04,nettelmann} to
estimate the mass of the present-day core, but it is unclear whether
these masses correspond to the primordial core mass. It has been
suggested \citep{guillot-book,saumon-apj-04} that the present-day core
mass of Jupiter may be too small to explain its formation by core
accretion within the relatively short lifetime of the protosolar
nebula \citep{pollack-icarus-96}, although a more recent Jupiter model
\citep{militzer-apj-08} predicted a larger core of 14--18 Earth masses
which is consistent with core accretion. Furthermore, direct
measurements of Jupiter's atmosphere suggest a significant enhancement
in the concentration of heavy $(Z > 3)$ elements
\citep{niemann-science-96}, but it is unknown to what extent this
should be attributed to a large flux of late-arriving planetesimals
versus the upwelling of core material. Determining the extent of core
erosion is thus a major priority for understanding the interiors of
giant planets and the process by which they were formed.

In this work we focus on water ice, presumed to be a major constituent
of the core, and consider the question of whether it has significant
solubility in fluid metallic hydrogen at conditions corresponding to
the core-mantle boundaries of giant gas planets. Water ice is the most
prevalent of the planetary ices (water, methane and ammonia) which may
be assumed to make up the outermost layers of a differentiated
rock-ice core \citep{hubbard-science-81}. At the conditions of
temperature and pressure prevalent at giant planet cores, water ice is
predicted \citep{cavazzoni,french-prb-09} to be in either in a fully
atomic fluid phase in which oxygen and hydrogen migrate freely and
independently, or in a superionic phase in which oxygen atoms vibrate
around defined lattice sites while hydrogen atoms migrate
freely. Assuming the existence of a core-mantle boundary at which
water ice and the fluid H-He phase are in direct contact, the relevant
question is the extent to which the system may lower its Gibbs free
energy by the redistribution of the atoms of the ice phase into the
fluid hydrogen. The extreme pressure and temperature conditions
prevalent at giant planet core-mantle boundaries (8000--12000K and
8--18 Mbar for Saturn, 18000--21000K and 35--45 Mbar for Jupiter) are
not yet obtainable in the laboratory, thus \emph{ab initio}
simulations provide the best available guide to determining the extent
of core solubility.

\section{Theory and Methodology}

We used density functional molecular dynamics (DFT-MD) calculations
and coupling constant integration (CCI) techniques to compute the
Gibbs free energy of solvation, $\Delta G_{sol}$, of H$_2$O in fluid
metallic hydrogen, i.e. the change in Gibbs free energy when an H$_2$O
molecule is removed from the pure ice phase and dissolved in H.  The
free energy of solubility is computed from the free energies of three
systems: pure ice, pure fluid H, and a mixed system in which the atoms
of one water molecule are dissolved in $n$ atoms of hydrogen,

\begin{equation}
\Delta G_{sol} = G \left(\mbox{O} \mbox{H}_{n+2}\right) - \left[G \left( {\mbox{H}_2 \mbox{O}} \right) + G\left( \mbox{H}_{n} \right) \right]
\label{deltag}
\end{equation}

where $G \left( {\mbox{H}_2 \mbox{O}} \right)$ is the energy per
$\mbox{H}_2\mbox{O}$ stoichiometric unit of the ice phase, and
$G\left(\mbox{H}_{n}\right)$ is obtained from an appropriately-scaled
simulation of 128 H atoms. This quantity becomes more negative as
solubility increases. A $\Delta G_{sol}$ of zero implies a saturation
concentration of exactly one H$_2$O to $n$ H. In order to span the
range of likely conditions for the core-mantle boundary of Jupiter and
Saturn, we considered pressures of 10, 20 and 40 Mbar, at a range of
temperatures from 2000 to 20000~K.

Computation of free energies from MD simulations is difficult since
the entropy term is not directly accessible. Here we use a two-step
CCI approach as previously applied by several authors
\citep{alfe-nature-99,morales-pnas-09,wilson-prl-10} to compute free
energies. The CCI method provides a general scheme for computing
$\Delta F$ between systems governed by potential energy functions
$U_1$ and $U_2$. We construct an artificial system $U_\lambda$ whose
forces are derived from a linear combination of the potential energies
of the two systems $U_\lambda = (1-\lambda)U_1 + \lambda U_2$. The
difference in Helmholtz free energy between the two systems is then

\begin{equation}
\Delta F = \int_0^1 \langle U_2 - U_1 \rangle_\lambda \: d \lambda,
\end{equation}

where the average is taken over the trajectories governed by the
potential $U_\lambda$.  We perform two CCIs for each $G$ calculation:
first from the DFT-MD system to a system governed by a classical pair
potential which we fit to the DFT dynamics of the system via a
force-matching approach \citep{izvekov-jcp-04}, and then from the
classical system to a reference system whose free energy is known
analytically.

\subsection{Material phases}

The material phases in question must be established prior to the Gibbs
free energy calculations. At the pressure and temperature conditions
of interest, hydrogen is a metallic fluid of H atoms in which
molecular bonds are not stable. Water dissolved within hydrogen
likewise is non-molecular, with a free O atom in an atomic H
fluid. For water ice, the phase diagram at giant planet core pressures
is divided into three regimes
\citep{cavazzoni,goldman,mattsson-prl-06,french-prb-09}: a
low-temperature ($<$~2000~K) crystalline regime, an intermediate
superionic regime in which oxygen atoms vibrate around fixed lattice
sites while hydrogen atoms migrate freely, and a higher-temperature
fully fluid regime in which both hydrogen and oxygen atoms are
mobile. The transition between the crystalline and superionic regimes
has not yet been studied in detail at high pressures but our
simulations find that it occurs below 2000~K. The transition from
superionic to fully fluid is found to occur in simulation at
temperatures ranging from 8000~K for 10 Mbar and 13000~K for 50 Mbar
\citep{french-prb-09}. Consequently our study includes the superionic
and fully fluid ice phases.

Previous studies of superionic ice \citep{cavazzoni,french-prb-09}
have used an \emph{bcc} arrangement of atoms for the oxygen
sublattice. We found that such a lattice was stable at all points
studied in the superionic regime except for 10 Mbar at 2000 and
3000~K. At these conditions we found that the oxygen sublattice was
stable in the \emph{Pbca} geometry which we recently reported to be
the most stable zero-temperature structure for ice at 10 Mbar
\citep{militzer-prl-10}. At 10 Mbar and 5000~K, superionic ice with a
\emph{bcc} sublattice was stable but the \emph{Pbca} was not. The
\emph{bcc} oxygen sublattice was found to be stable for 20 and 40 Mbar
pressures at all temperatures studied in the superionic regime.
Attempts to perform a superionic simulation in the \emph{Cmcm}
geometry reported by \cite{militzer-prl-10} to be the stable
zero-temperature structure at these pressures resulted in an unstable
system with a large anisotropic strain. We thus used a \emph{bcc}
oxygen sublattice for all superionic ice simulations, except the
2000~K and 3000~K simulations at 10 Mbar which used the oxygen
sublattice from the \emph{Pbca} phase, as indicated in Figure 1. While
we cannot yet exclude the possibility of the existence of yet another
superionic ice structure, we note that the differences between
different ice phase energies are found to be on the order 0.1~eV per
H$_2$O, and thus will not significantly affect our results about the
stability of ice in giant planet cores.

\subsection{Computation of Gibbs free energies}

Our goal in this paper is to study the solubility of water ice in
fluid hydrogen. Due to considerations arising from the entropy of
mixing, the solubility of one material in another is never zero,
however solubility at trace quantities is not sufficient for core
erosion. In particular, we wish to know whether solubility is
thermodynamically favored at concentrations significantly greater
than the background concentration of oxygen in the fluid envelope of
Jupiter or Saturn -- this is equal to approximately one O atom to 1000
H atoms if we assume solar concentrations for the Jovian envelope and
approximately one part in 300 if we assume a threefold enrichment for
oxygen as observed for most other heavy elements \citep{mahaffy}. We
begin by computing the Gibbs free energies of solubility for
dissolving H$_2$O in pure H at one part in 125, and generalize later.

The coupling constant integration approach requires, as an integration
end point, a reference system whose free energy may be computed
analytically. It is important to ensure that the system does not
undergo a phase change along the integration pathway since this may
cause numerical difficulties in the integration. For the fluid
systems, being pure hydrogen, hydrogen with oxygen, and ice at
temperatures above the superionic-to-fluid transition, we used an
ideal atomic gas as the reference system
\citep{alfe-nature-99,morales-pnas-09,wilson-prl-10}. For superionic
ice, we use as a reference system a combination of an ideal gas system
for the hydrogen atoms with an Einstein crystal of oxygen atoms each
oxygen tethered to its ideal lattice site with a harmonic potential of
spring constant 30~eV/$\mbox{\AA}^2$.

The DFT-MD simulations in this work used the Vienna Ab Initio
Simulation Package (VASP) \citep{vasp} with pseudopotentials of the
projector augmented wave type \citep{paw} and the exchange-correlation
functional of \citet{pbe}. The pseudopotentials used had a core radius of 0.8~{\AA} for hydrogen and 1.1~{\AA} for oxygen. Wavefunctions were expanded in a basis set
of plane waves with a 900~eV cutoff and the
Brillouin Zone was sampled with $2 \times 2 \times 2$ $k$-points. The
electronic temperature effects were taken into account via Fermi-Dirac
smearing. A new set of force-matched potentials was fitted for each
pressure-temperature conditions for each stoichiometry. All MD
simulations used a 0.2~fs timestep. In the classical potential under
superionic conditions, an additional harmonic potential term was added
to ensure that oxygen atoms remained in the appropriate lattice sites,
however, we found that in most cases the fitted pair potential alone
was sufficient to stabilize the superionic state.

The first step of the free energy calculations was the determination
of the appropriate supercell volumes for each system for each set of
pressure and temperature conditions. This was accomplished via
constant-pressure MD simulations
\citep{hernandez-jcp-01,hernandez-prl-10} with a duration of 1.6~ps
(0.6~ps for ice). DFT-MD trajectories were then computed in a fixed
cell geometry in order to fit classical potentials. A run of 0.4~ps
was found to be sufficient for fitting suitable potential. We then
performed molecular dynamics runs 600 fs long (400 fs for ice) at five
$\lambda$ values to integrate between the DFT and classical
systems. Finally, we performed classical Metropolis Monte Carlo at 24
$\lambda$ values to integrate from the classical to the reference
system.

\section{Simulation results}

Table I lists the total Gibbs free energies for each simulated system
(H$_{128}$, OH$_{127}$ and ice) for each set of temperature and
pressure conditions. Ice was confirmed to remain superionic except at
the 20~Mbar/12000~K and 40~Mbar/20000~K conditions where it was fully
fluid. The error bars on the computed $G$ values are dominated primarily by the uncertainty in the computed volume at the desired pressure, and secondarily by uncertainty in the $\langle U_{DFT} - U_{classical} \rangle_\lambda$ term in the coupling constant integration. These free energies are combined using Equation \ref{deltag} to
give $\Delta G_{sol}$ representing the free energy change associated
with removing an H$_2$O from the ice phase and dissolving it in the
hydrogen fluid at a concentration of molecule per 125 solute H
atoms. $\Delta G_{sol}$ increases strongly with temperature, but shows
only a weak dependence on pressure within the 10--40 Mbar range under
consideration. The $\Delta G$ values were found to be well converged with respect to wavefunction cutoff, k-point sampling to within the available error bars. The effect of the electronic entropy term on $\Delta
G_{sol}$ was found to be less than 0.1 eV in all cases.  From a linear
interpolation through the adjacent data points we estimated the
temperature at which $\Delta G_{sol}$ passes through zero as 2400~K
$\pm$ 200~K at 40 Mbar, 2800~K $\pm$ 200~K at 20 Mbar, and 3400~K
$\pm$ 600~K at 10 Mbar.  As shown in Figure 1, this is clearly far
lower than any reasonable estimate for the core-mantle boundaries for
either Jupiter or Saturn. The onset of high solubility occurs within
the portion of the phase diagram where ice is superionic, and does not
coincide with either the superionic-to-fluid transition or the
crystalline-to-superionic transition in ice.

The total $\Delta G$ of solubility may be broken down into three
components: a $\Delta U$ term from the potential energy, a $P \Delta
V$ term from the volume difference, and a $-T \Delta S$ entropic
term. Figure 2 shows this breakdown as a function of temperature for
20~Mbar.  The breakdown for other pressures looks similar. The $\Delta
U$ term, representing the difference in chemical binding energy,
provides approximately a 2 eV per molecule preference for ice
formation at all temperatures where ice is superionic, but a much
smaller preference for the 12000~K case where ice is fluid.  The $PV$
term is indistinguishable from zero within the error bars, suggesting
that this is not a pressure-driven transition, in contrast to recent
results on the partitioning of noble gases between hydrogen and helium
in giant planet interiors in which volume effects were found to be the
dominant term \citep{wilson-prl-10}. The $-T \Delta S$ term dominates
the temperature-dependent behavior, underlining that ice dissolution
is indeed an entropy-driven process.

Given the Gibbs free energy of solubility for the insertion of one
H$_2$O into 125 H atoms we can determine an approximate Gibbs free
energy of solubility at other concentrations, by neglecting the
contribution of the oxygen-oxygen interaction and including only the
entropic term arising from the mixing.  Under these approximations, we
obtain the expression

\begin{eqnarray}
\frac{\Delta G_{sol}[m] - \Delta G_{sol}[n]}{k_BT} &=& (m+2) \log \left( \frac{(m+2)V_H + V_O}{(m+2)V_H}\right)\nonumber \\ 
&-& (n+2) \log \left( \frac{(n+2)V_H + V_O}{(n+2)V_H} \right)\nonumber \\
&-& \log \left( \frac{m+2}{n+2} \right) ,
\label{conc}
\end{eqnarray}

where $V_H$ and $V_O$ are the effective volumes of each H and O atom
in the fluid. This approximation becomes invalid as the oxygen-oxygen
interaction term in the fluid oxygen phase becomes significant,
however for our purposes it is sufficient to know that the saturation
concentration is significantly higher than the background
concentration. If a saturation concentration of oxygen in hydrogen
does indeed exist then this may be expected to have a retarding effect
on the erosion of the core.

We must also consider possibility that hydrogen and oxygen could
dissolve separately from the H$_2$O mixture, leaving behind a
condensed phase with stoichiometry other than H$_2$O. We tested two
cases explicitly, computing the free energies of pure oxygen and
one-to-one HO phases at the Jupiter-like 40 Mbar, 20,000~K set of
conditions. Both O and HO were found to be in a fully fluid state at these temperature/pressure conditions. We found that HO had a free energy of solubility of -11.2
eV per oxygen, while pure oxygen had -22.8 eV per oxygen. Comparing to
the -8.9 eV per oxygen solubility of H$_2$O, this suggests oxygen-rich
condensed phases are less thermodynamically stable than the H$_2$O
phase, and certainly far less favourable than dissolution of the dense
phase into metallic hydrogen. We have also neglected the possibility
of hydrogen-enriched dense phases such as H$_3$O. While it possible
such phases may be somewhat energetically preferred to H$_2$O, it is
extremely unlikely that the preference will be strong compared to the
$8-12$ eV per O unit preference for solubility.

In Figure 3 we plot the estimated $\Delta G_{sol}$ as a function of
concentration for various computed temperatures at a pressure of
40~Mbar. A negative value of $\Delta G_{sol}$ means that it is
thermodynamically favorable to dissolve further ice into the hydrogen
phase at a given hydrogen-phase ice concentration, and the point at
which $\Delta G_{sol}$ is zero is the saturation concentration.  For
2000~K and 3000~K the saturation concentrations are estimated to be on
the order of 1:500 and 1:20 respectively, while for higher
temperatures the saturation concentration is much higher. Given that
we neglect O--O interactions in the fluid hydrogen phase, it is
difficult to precisely determine the saturation concentrations for
higher temperatures. However, we may safely say that the saturation
concentration for temperatures in excess of 3000~K is very much
greater than the background oxygen concentration, and hence that
solubility of ice into hydrogen at the core-mantle boundaries of
Jupiter and Saturn is expected to be strongly thermodynamically
favored.

\section{Discussion}

The consequences of core erosion for planetary evolution models have
been previously considered by \citet{stevenson-pss-82} and later
\citet{guillot-book}. The effects of core erosion can potentially be
detected either by orbital probes such as Juno or by atmospheric entry
probes, since the redistribution of core material throughout the
planet will manifest itself both by a smaller core (detectable from
gravitational moments) and a higher concentration of heavy elements in
the atmosphere than would be expected in a planet without core
erosion. Once core material has dissolved into the metallic H layers,
the rate at which core material can be redistributed throughout the
planet is expected to be limited by double diffusive convection
\citep{turner,huppert}. Since the higher density due to compositional
gradients of the lower material interferes with the convection
process, convection may be slowed significantly. \citet{guillot-book}
modelled the effect of core erosion under the assumptions of fully
soluble 30 Mbar core in each planet. Under their assumptions up to 19
Earth masses could have been redistributed from Jupiter's core but
only 2 Earth masses from Saturn's, the difference being Jupiter's
higher temperatures. While this prediction is subject to significant
uncertainty in many aspects of the model, it does suggest that a
redistribution of a significant fraction of the initial protocore is
possible, at least in Jupiter.  Further refinement of models for the
upconvection of core material and its observable consequences may be
fruitful. The effect of core erosion on the heat transport and mass
distribution properties of Jupiter and Saturn should also be taken
into account in future static models of these planets' interiors.

These calculations can be expanded in several ways. We have neglected
the presence of helium in the hydrogen-rich mantle, however due to the
large magnitude of $\Delta G_{sol}$ and the chemical inertness of
helium we do not expect the presence of helium in the mantle to
significantly affect the solubility behavior. We have explicitly made
the assumption that ice and hydrogen are in direct contact, an
assumption which might fail in one of two ways. If hydrogen and helium
are immiscible at the base of the atmosphere then the core may make
direct contact with a helium-rich layer. This, however, is unlikely in
the context of the calculations of \citet{morales-pnas-09} who predict
hydrogen-helium immiscibility only far away from the cores of Jupiter
and Saturn.  The other possibility is that the ice layers of the core
may be gravitationally differentiated, leaving ice beneath layers of
the less dense planetary ices methane and ammonia. Since the bonding
in these is similar to that in water ice one could assume that they
show a similar solubility behavior, but this analysis is the subject
of future work.

We have also considered only the structure of the present-day
planet. As suggested by \citet{slav}, a proper consideration of core
solubility must also include solubility during the formation process,
as dissolution of the icy parts of the core into the accreting
hydrogen during the formation may result in the amount of ice on the
core itself being small by the time the planet reaches its final
size. A treatment of the formation processes for Jupiter and Saturn
using ice solubilities derived from \emph{ab initio} calculations may
be valuable.

Our calculations strongly suggest that icy core components are highly
soluble in the fluid mantle under the conditions prevalent at the
core-mantle boundaries of Jupiter and Saturn. Since many
recently-discovered exoplanets are more massive and hence internally
hotter than Jupiter, it can be expected that any initial icy cores in
these exoplanets will also dissolve. The presence of core erosion may allow models predicting a small present-day Jovian core to be made consistent with the large initial core required by core erosion, however models of the interior mass distribution of the planet will need to be revised to take the inhomogeneous composition of the lower layers implied by convection-limited core redistribution into account. Improved models which
include core redistribution processes, combined with the
data from the Juno probe, may assist in understanding the history and
present structure of Jupiter and other planets in our own and in other
solar systems.

%%%%%%%%%%%%%%%%%%%%%%%%%%%%%%%%%%%%%%%%%%%%%%%%%%%%%%%%%%%%%%%%%%%%

\acknowledgments

 This work was supported by NASA and NSF. Computational resources were supplied in part by TAC, NCCS and NERSC. We thank D. Stevenson for discussions.

%\bibliographystyle{apj}
%\bibliography{jupiter}

\clearpage

\bibliographystyle{apj}

\clearpage

\begin{table}
\begin{tabular} {c c |c c c c }
\hline
P & T& $G$(H$_{128}$) & $G$(H$_{127}$O) & $G$(Ice) & $\Delta G_{sol}$ \\
(Mbar) &(K) & (eV) & (eV) & (eV) & (eV) \\
\hline
\hline
10 & 2000  & 716.1(5)& 728.4(4) &1800.2(22)$^*$ &0.97(62) \\
10 & 3000  & 672.2(3)& 683.9(4) &1731.4(22)$^*$ &0.35(52) \\
10 & 5000  & 561.8(3)&571.7(3) &1311.5(11)& -1.10(50) \\
20 & 2000  & 1308.3(1)&1332.8(2) &2945.3(10) & 0.61(24) \\
20 & 3000  & 1271.2(1)&1295.1(2) &2895.8(9) & 0.11(22)\\
20 & 5000  & 1173.0(3)&1195.3(2) &2795.2(15) &-1.91(37) \\
20 & 8000  & 989.7(7)& 1009.9(4)& 2536.5(15)& -3.57(80)\\
20 & 12000 & 705.9(10)&722.5(6) &2049.8(26)$^\dagger$ & -4.8(12)\\
40 & 2000  & 2161.5(1)&2205.5(1) &5102.2(10) &0.18(14) \\
40 & 3000  & 2131.3(1) &2174.8(1) &5062.6(10) &-0.25(17) \\
40 & 5000  & 2046.0(3)& 2087.9(2)& 4943.0(13)& -1.69(30)\\
40 & 8000  & 1880.8(4)& 1920.3(2)& 4701.3(13)& -3.47(43)\\
40 & 20000 & 1011.6(17)&1041.1(14) &3350.6(30)$^\dagger$ &-8.9(22) \\
\hline
\end{tabular}
\caption{Gibbs free energies per simulation cell of fluid hydrogen,
  fluid hydrogen with oxygen, and ice at each temperature/pressure
  combination, along with the energy of solvation $\Delta G_{sol}$ per
  H$_2$O molecule. Ice systems marked with a star have
  H$_{128}$O$_{64}$ stoichiometry and an oxygen structure from the
  \emph{Pbca} phase while other ice systems have H$_{108}$O$_{54}$ and
  a \emph{bcc} oxygen sublattice. Ice systems marked with a dagger are
  fully fluid, while others are superionic.}
\label{table3}
\end{table}

\clearpage

\begin{figure}[t]
\includegraphics{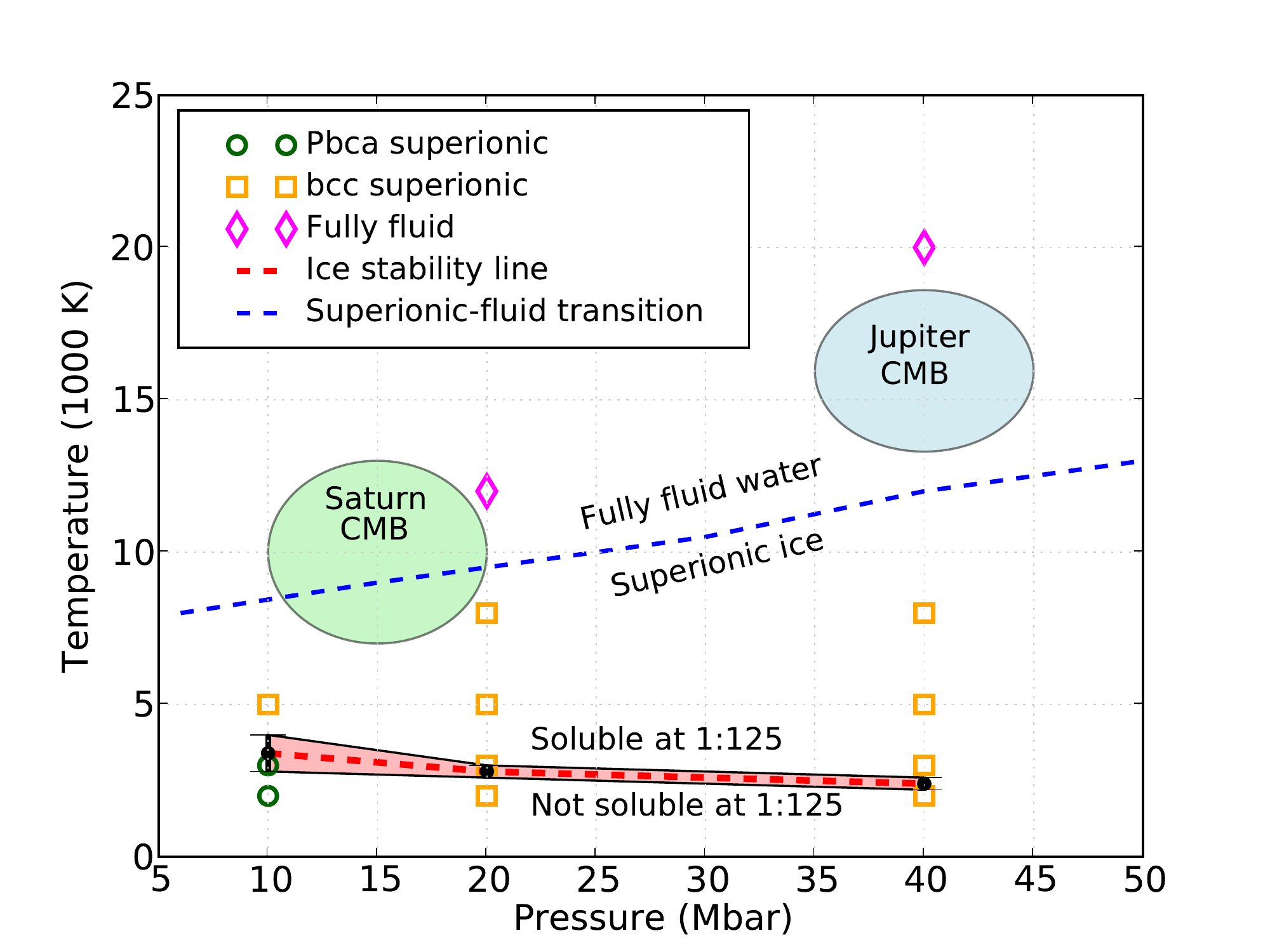} 
\caption{(Color online) Computed location of the temperature at which
  saturation is reached at 1:125, as a function of pressure between 10
  and 40~Mbar, with the shaded area corresponding to error bars, in
  relation to the approximate location of the superionic-to-fluid
  transition \citep{french-prb-09}, and the approximate conditions
  corresponding to the core-mantle boundaries of Jupiter and
  Saturn. The points at which calculations have been undertaken are
  marked, with the shape and color of the symbols indicating
  phase of ice used in each calculation.}
\label{phase_diagram}
\end{figure}

\clearpage

\begin{figure}[t]
\includegraphics{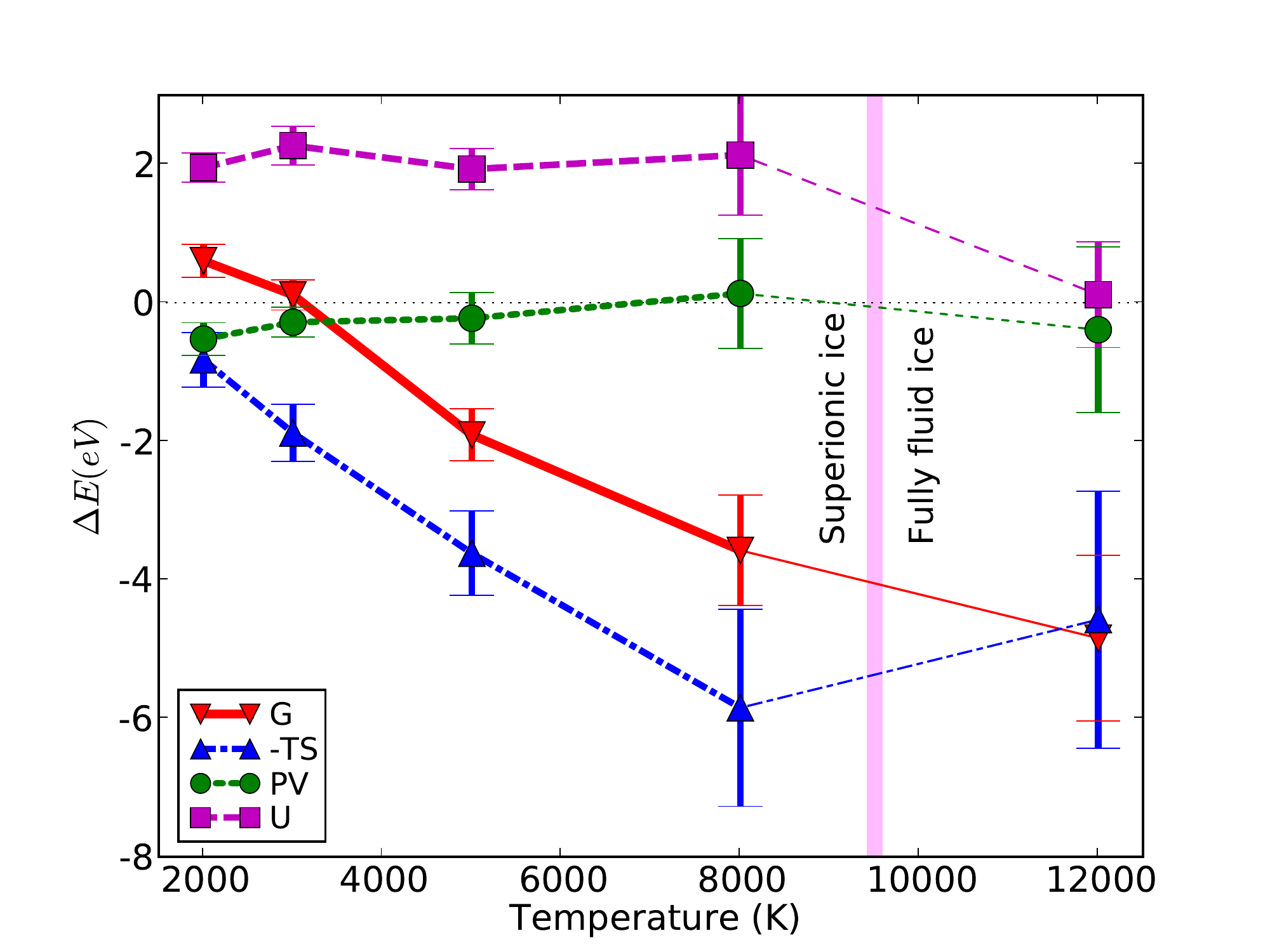}
\caption{Breakdown of the computed $\Delta G_{sol}$ into contributions
  from the $\Delta U$, $P \Delta V$, and $-T \Delta S$ terms, as a
  function of temperature, at a pressure of 20 Mbar.}
\label{phase_diagram}
\end{figure}

\clearpage

\begin{figure}[t]
\includegraphics{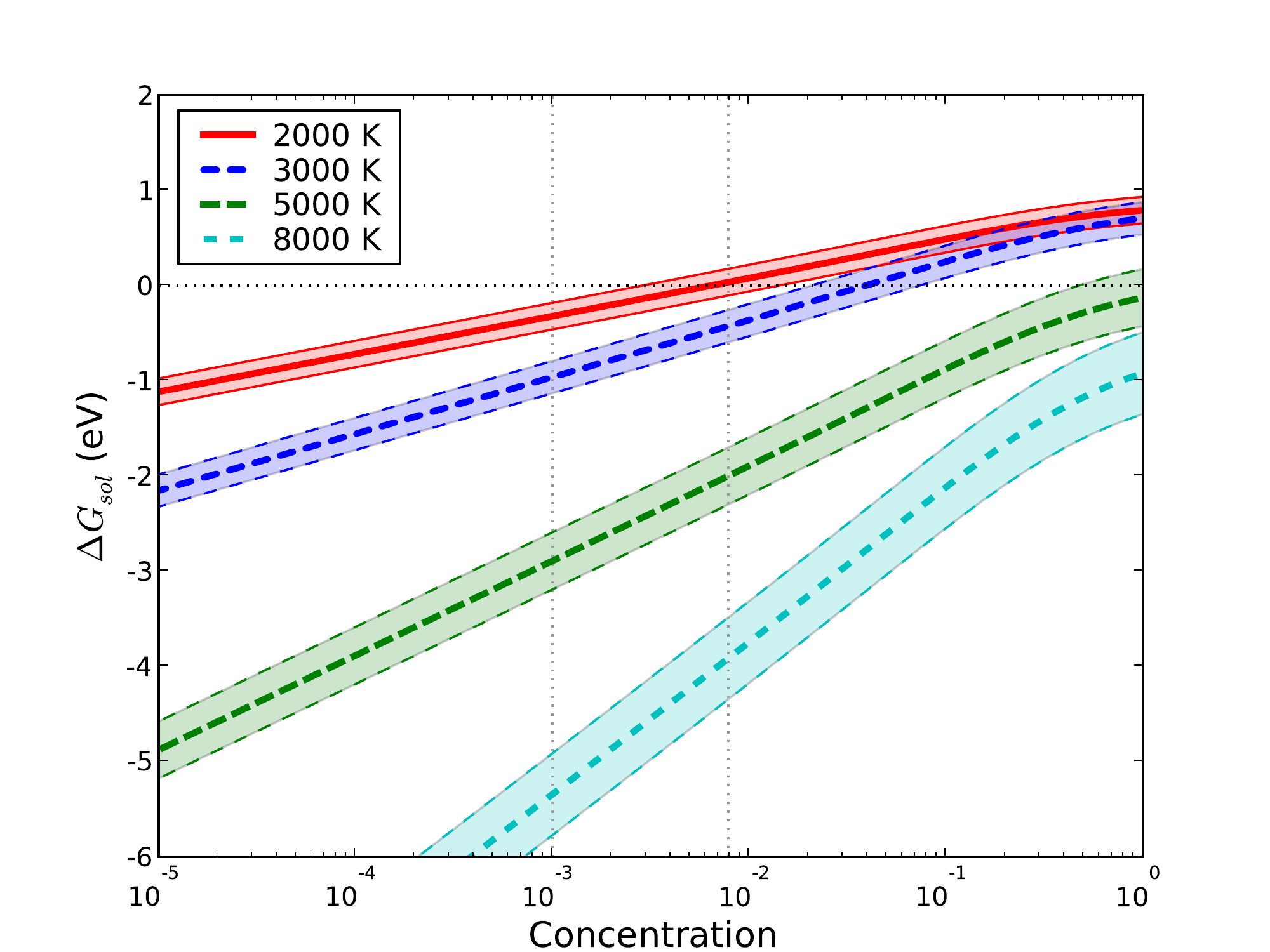} 
\caption{Estimated free energy of solubility $\Delta G_{sol}$ of ice
  in hydrogen as a function of oxygen concentration in the limit where
  oxygen atoms do not interact in the fluid phase, for various
  temperatures at a pressure of 40 Mbar. Shaded regions indicate error
  bars.}
\label{phase_diagram}
\end{figure}

\end{document}